# ELECTRON CLOUD EFFECTS: CODES AND SIMULATIONS AT KEK

K. Ohmi, KEK, 1-1 Oho, Tsukuba, 305-0801, Japan

*Abstract*

Electron cloud effects had been studied at KEK-Photon Factory since 1995. e-p instability had been studied in proton rings since 1965 in BINP, ISR and PSR. Study of electron cloud effects with the present style, which was based on numerical simulations, started at 1995 in positron storage rings. The instability observed in KEK-PF gave a strong impact to B factories, KEKB and PEP-II, which were final stage of their design in those days. History of cure for electron cloud instability overlapped the progress of luminosity performance in KEKB. The studies on electron cloud codes and simulations in KEK are presented.

## INTRODUCTION

Electron cloud instability in positron storage ring had been observed at KEK-PF since start of positron operation in 1988. It had not been identified as an electron cloud effect. KEK-PF had suffered very strong vertical coupled bunch instability. Design of KEKB completed 1994-1995. The positron beam instability had to be solved to complete KEKB Low Energy Ring (LER) design. M. Izawa, Y. Sato and T. Toyomasu had performed many experiments [1] and studied a model to solve it. They showed that a short-range wake gave observed mode spectra. They studied a model, in which electrons trapping by beam under the condition of electron-ion plasma [2].

K. Ohmi had studied a possible model to explain the instability. Photoelectrons, which are supplied continuously (every passage of bunches) from the chamber wall, can induce strong coupled-bunch instability [3]. The simulation method is transferred to SLAC-LBNL team to study in PEP-II LER immediately. The instability was confirmed by a series of experiments at BEPC in IHEP, China [4], which was collaboration of IHEP and KEK.

The electron cloud effects have been studied vigorously in the B factories, KEKB and PEP-II, and then LHC in 1990's. Operation of B factories started in 1998-1999. The instability has been observed in KEKB-LER. Weak solenoid coils were wound along the whole ring to protect electrons near the beam. The unstable mode of the coupled bunch instability was clearly related to the solenoid status, ON or OFF [5,6]. Corrective electron motion reflected to beam unstable mode.

Single bunch instability has been observed in KEKB-LER [7,8]. Generally fast head-tail instability is caused by merge between 0 and -1 synchrotron sideband modes in positron ring. In the single bunch instability, clear positive side band $\nu_y+a\nu_s$ (1<a<2) has been observed [9].

Simulations have been performed to explain both single and coupled bunch instabilities. We present the simulation results with focusing unstable mode in this paper.

## COUPLED BUNCH INSTABILITY DUE TO ELECTRON CLOUD

### Simulation of electron cloud build up and coupled bunch instability

Beam-electron cloud system in multi-bunch regime is described by following equations:

$$\frac{d^2\boldsymbol{x}_p}{ds^2} + K(s)\boldsymbol{x}_p = \frac{2N_e r_e}{\gamma}\sum_{e=1}^{N_e} \boldsymbol{F}_G(\boldsymbol{x}_p - \boldsymbol{x}_e)\delta_P(s-s_e) \quad (1)$$

$$\frac{d^2\boldsymbol{x}_e}{dt^2} = \frac{e}{m_e}\frac{d\boldsymbol{x}_e}{dt}\times\boldsymbol{B} - 2N_p r_e c\sum_{p=1}^{N_p} \boldsymbol{F}_G(\boldsymbol{x}_e-\boldsymbol{x}_p)\delta_P(t-t_p(s_e)) - r_e c^2\frac{\partial\phi(\boldsymbol{x}_e)}{\partial \boldsymbol{x}_e}, \quad (2)$$

where the electric potential of electron cloud is given by

$$\triangle\phi(\boldsymbol{x}) = \sum_{e=1}^{N_e}\delta(\boldsymbol{x}-\boldsymbol{x}_e) \quad . \quad (3)$$

Bunches, which are rigid Gaussian shape in transverse and are located equal spacing along s, are represented by their center of mass $\boldsymbol{x}_p$. $\boldsymbol{F}_G$ is expressed by the Bassetti-Erskine formula.

The electron cloud build-up is simulated by integrating the second equation for the motion of macro-electrons $\boldsymbol{x}_e$ under $\boldsymbol{x}_p$=0. The initial condition of electrons, where and when electrons are created, is sketched in Figure 1. When beam pass through the chamber cross-section, electrons are created with an energy distribution. When an electron hits the wall, secondary electrons are produced with a probability.

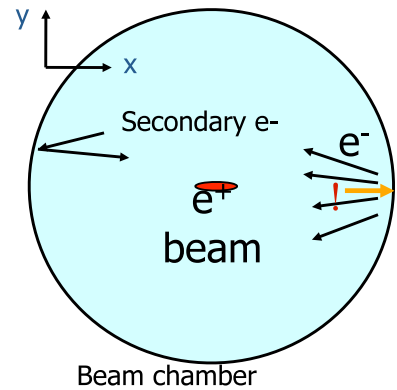

Figure 1: Electron cloud build-up model.

The coupled bunch instability is studied by solving the two coupled equations (1) and (2).

### Coupled bunch instability in KEK-PF

Progress of electron cloud effects in these 17 years (since 1995) started from the interpretation of an instability observed in KEK-Photon Factory. Very strong coupled-bunch instability had been observed since the positron storage had started. The threshold of the instability is very low 10-15mA. Either of positron or electron could be stored by changing the polarity of the magnets. The instability is observed only in positron storage. Figure 2 shows frequency spectrum for the instability published in [1]. The first electron cloud build-up code (PEI) is developed in 1995 [3], and the coupled bunch instability was interpreted as a wake effect of the electron cloud. The wake force was estimated by perturbation of the cloud due to a passage through of a shifted bunch [3]. Figure 3 shows the simulated wake force. The shape of wake force contains characteristic of the electron motion in the drift space. The wake force is the same direction as the shift of the bunch: namely, the bunch shift upper induces the wake force, by which following bunches near the shifted bunch are kicked upper. Namely, an upper shift of bunch, which induces electron motion toward upper direction, results upper force for following bunches. This feature is explained in Figure 4. This wake has the regular property as seen in [10]. Note the definition of the sign of the wake force is different.

Growth rate for each mode is calculated by the wake force as follows,

$$(\Omega_m - \omega_\beta)L/c = \frac{N_p r_e c}{2\gamma\omega_\beta} \sum W_n \exp\left(2\pi i n \frac{m+\nu_\beta}{h}\right) \quad (4)$$

The positive imaginary part of the frequency ($\Omega_m$) is the growth rate for m-th mode. Figure 5 shows unstable mode spectrum estimated by the simulated wake force. The growth rate is very fast, 0.3msec for m=250-300. The wake force ($\omega<h\omega_0$) with the regular property induces higher mode near the beam harmonic number. The corresponding frequency is (h-m-$\nu_y$)$f_0$, where h and $f_0$ are beam harmonic number and revolution frequency, respectively. The growth rate and mode spectrum well agreed with the measurements.

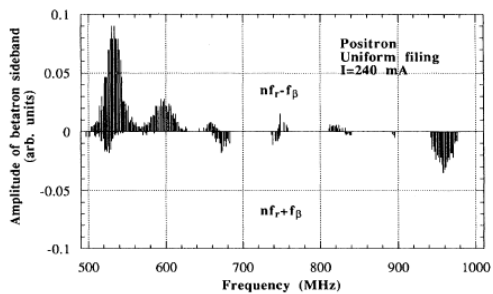

Figure 2: Unstable mode spectrum measured in KEK-PF [1].

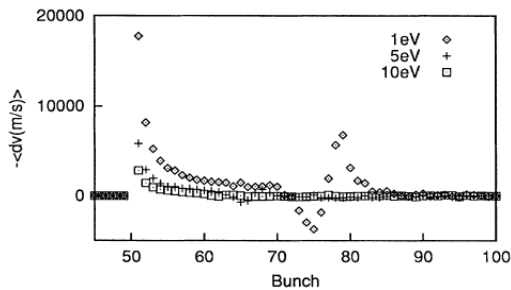

Figure 3: Simulated wake force due to electron cloud in KEK-PF [3].

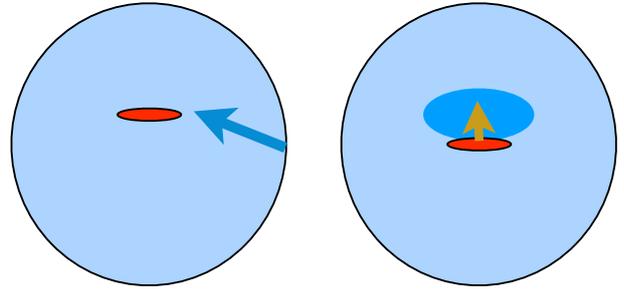

Figure 4: Wake force appearance due to electron cloud in drift space

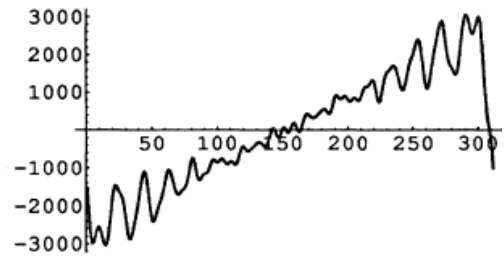

Figure 5: Unstable mode spectra due to the wake force in Figure 3 [3].

## Coupled bunch instability in BEPC

Experiments were held to study reappearance of the electron cloud instability in BEPC since 1997. Both/either of positron and/or electron beams could be stored in BEPC; where the circulating directions are opposite.

Figure 6 shows mode spectra in positron and electron storage, respectively. Multi-mode instability similar as KEK-PF observation (Figure 2) was seen in BEPC. While single mode instability in electron storage also agreed with KEK-PF observation.

Simulation using Eqs.(1) and (2) gives beam centroid motion under the beam-electron cloud interactions. FFT of the centroid positions gives unstable mode spectrum, which is compared with measurement. The spectra well agree with the measurements as shown in Figure 7. These features of unstable modes in both of electron and positron instability are induced by the wake force with the regular property [10] as shown in Figure 3.

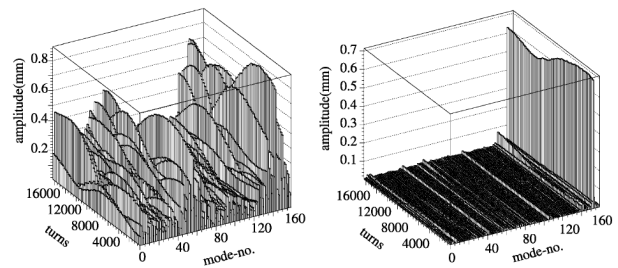

Figure 6: Mode spectra for electron cloud and ion instabilities, respectively [4]. Left and right pictures show mode variation in positron and electron storage, respectively.

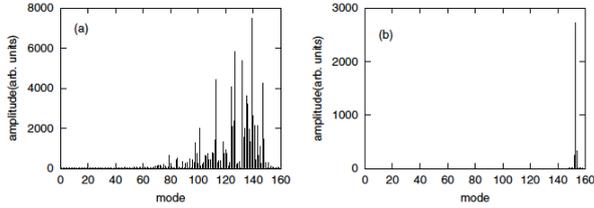

Figure 7: Mode spectra given by simulation for electron cloud and ion instabilities, respectively [4].

*Coupled bunch instability in KEKB-LER*

Coupled bunch instability due to electron cloud has been observed since the early stage of the commissioning of KEKB. For narrow bunch spacing 4-6 ns, fast beam losses was serious in the operation.

Solenoid coils were wound at the magnet free section to protect the electrons coming to the beam position. The field strength is ~<50G. The solenoid coils are covered 95% of the magnet free section finally. The solenoid coil did not work well to suppress the coupled bunch instability, while it works to suppress the single bunch instability very well. Electrons stay longer time in the vacuum chamber than that for the case of no solenoid. R/Q of the wake force induced by electron cloud was reduced, but the range of the wake field (Q) was longer. Bunch-by-bunch feedback system suppressed the coupled bunch instability for wider spacing >6ns.

Systematic measurement and simulations in KEKB were published in [5,6], respectively. The electron motion in a weak solenoid field (~<50 G) differs from the one in a drift space. Electrons rotate along the chamber wall and do not approach to the beam. The electron motion reflects the coupled bunch instability signal.

Figure 8 shows the horizontal mode spectrum for solenoid OFF: that is, in drift space. The left picture was given by measurement. The right picture was given by simulation for electron cloud in drift space. Vertical spectra for measurement and simulation were similar as horizontal ones in Figure 8. The feature of unstable modes is induced by the wake force with the regular property [10] as shown in Figure 3. The spectrum had very good agreement with simulations. Figure 9 shows two cuts of movie for simulated beam-electron motion in drift space. Coherent motion between bunches and electron flow is seen.

There is a peak of unstable modes near m=200 in both measurement and simulation. The frequency is 24 MHz (m+$\nu_y$) or 100MHz (h-m-$\nu_y$).

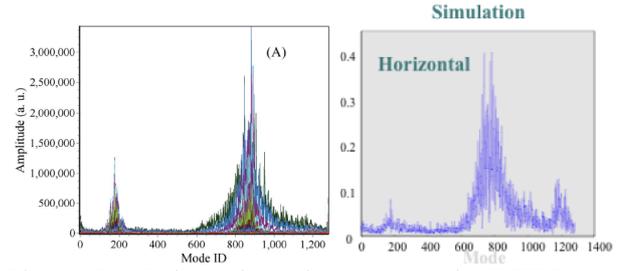

Figure 8: Horizontal mode spectrum in KEKB. Left picture is given by measurement with solenoid OFF [5,6]. Right picture is simulated for electron cloud in a drift space.

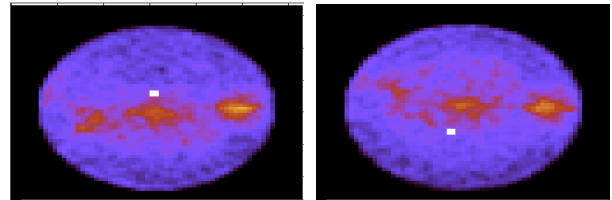

Figure 9: Two cut of movie of the beam electron motion. The white point indicates the beam position pass through the chamber.

Figure 10 shows the horizontal mode spectrum for solenoid ON. The left and right pictures are given for measurement and simulation. Vertical spectra for measurement and simulation were similar as Figure 10. The spectrum had very good agreement with simulations. Unstable mode was measured for various $B_z$ as shown in Figure 11.

Figure 12 shows two cuts of movie for simulated beam-electron motion in solenoid magnet $B_z$=10G. Coherent motion between bunches and electron rotation along the chamber is seen.

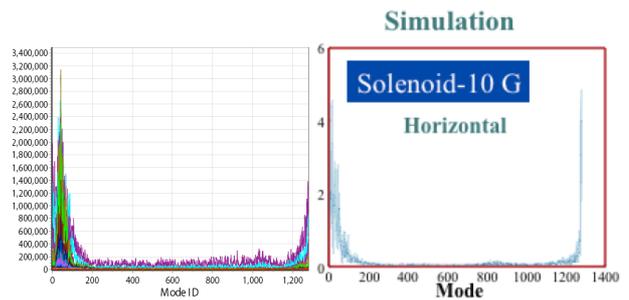

Figure 10: Horizontal mode spectrum in KEKB. Left picture is given by measurement with solenoid ON [5,6]. Right picture is simulated by electron cloud in solenoid field 10G.

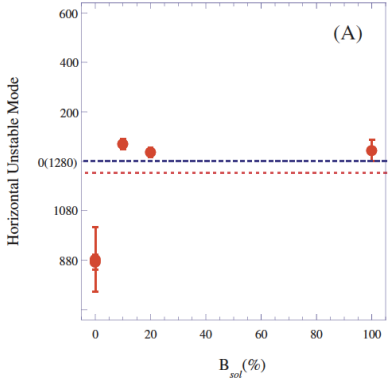

Figure 11: Unstable mode as a function of the solenoid strength [5]. 100% means $B_z$=~50G.

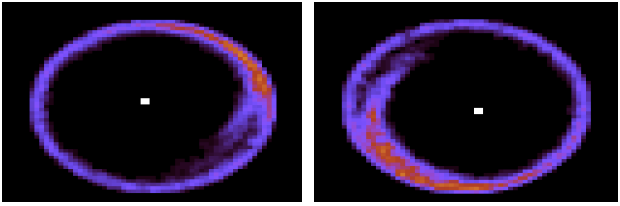

Figure 12: Two cut of movie of beam-electron motion in solenoid magnet.

Electrons moving in the solenoid field with a central force have two frequencies as follows,

$$\omega_\pm = \frac{\omega_c}{2} \pm \sqrt{\frac{\omega_c^2}{4} - \frac{r_e \bar{\lambda}_p c^2}{\bar{r}^2}}$$

Figure 13 shows the two types of frequencies. The chamber radius of KEKB is 5 cm. The upper and lower frequencies correspond to the cyclotron motion and rotation along chamber surface. The measurement showed lower solenoid field induced higher frequency mode as shown in Figure 11; m=70, f=(m+$\nu_y$)$f_0$=11MHz or (h-m-$\nu_y$)$f_0$=117MHz for 5 G, and m=30, f=7MHz or 121MHz for 10 G. The lower frequency $\omega_-$ agrees with lower betatron sideband (m+$\nu_y$)$f_0$. The wake force with the regular property [10] does not induce the lower sideband. The wake force is opposite direction (irregular) for the shift of the previous bunch. Figure 14 shows the wake force due to electron cloud in solenoid field. The irregular property is explained in Figure 15. A bunch shifted vertically induced electrons from a side, left or right depending on the solenoid polarity. The electrons rotate along the vacuum chamber. The wake force has a negative (opposite from the shift) peak after $\pi/2$ rotation.

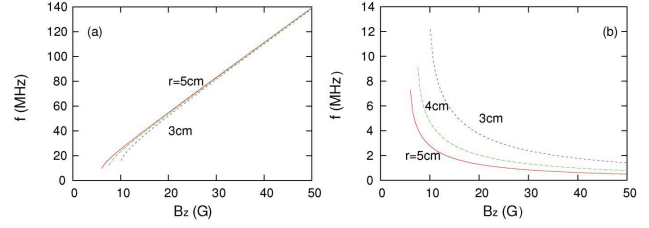

Figure 13: Magnetron frequency as a function of solenoid magnet strength $B_z$ [6]. Plots (a) and (b) depicts frequencies $\omega_+$ (~$\omega_c$, cyclotron mode) and $\omega_-$ (slow rotational mode), respectively.

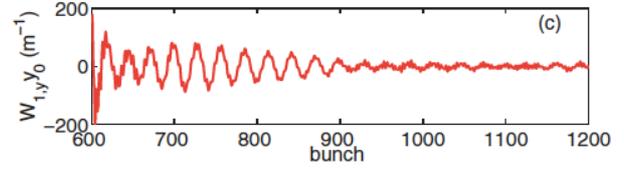

Figure 14: Wake force due to electron cloud in solenoid field [6].

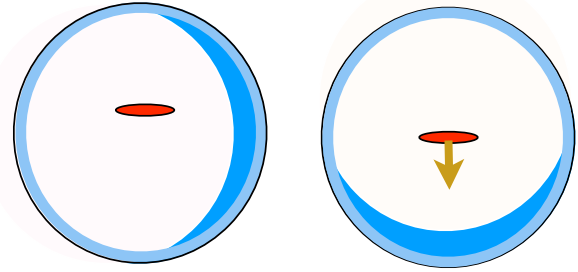

Figure 15: Irregular property of the wake force. Electrons induced by a shifted beam and their motion in solenoid field. The wake force is opposite to the shift.

## SINGLE BUNCH INSTABILITY DUE TO ELECTRON CLOUD

Single bunch fast head-tail instability caused by electron cloud has been observed in KEKB [7,8,9]. When beam current exceeds a threshold value, emittance increases and synchro-beta side band signal has been observed.

The simulation of the fast head-tail instability is performed by solving following equations [8]

$$\frac{d^2 \boldsymbol{x}_p}{ds^2} + K(s)\boldsymbol{x}_p = \frac{r_e}{\gamma}\frac{\partial \phi_e(\boldsymbol{x}_p)}{\partial \boldsymbol{x}_p}\delta_P(s - s_e)$$
$$\frac{d^2 \boldsymbol{x}_e}{dt^2} = \frac{e}{m_e}\frac{d\boldsymbol{x}_e}{dt} \times \boldsymbol{B} - r_e c^2 \frac{\partial \phi_p(\boldsymbol{x}_e)}{\partial \boldsymbol{x}_e}\delta_P(t - t_p(s_e))$$
(6)

Each potential of beam and electrons are solved using PIC algorithm. Since beam (1mmx0.1mm) is localized at the chamber center, free boundary condition is employed. Electron cloud is initialized every interactions with beam with a flat distribution ~40$\sigma_x$x60$\sigma_y$.

The single bunch instability signal has also been observed in PETRA-III and Cesr-TA, recently.

*Single bunch instability in KEKB-LER*

A beam size blow-up had been observed since early stage of KEKB operation around ~1999 [7]. The blow-up limited the luminosity performance. Figure 16 shows beam size blow-up and luminosity limitation in 2000-2001. Solenoid coils were wound 2000-2001 in 50%-70% of the whole drift space. Left picture shows beam size blowup without (green) and with (red) the weak solenoid field. Threshold of the beam size blow-up increases from 400 to 800mA. Left picture shows luminosity as function of current. Luminosity was saturated around 550mA at 2000 December. After winding solenoid additionally, 50% to 70%, luminosity was not saturated by 700mA at 2001 March. The solenoid coil was wound further after 2001, and covered 95% of drift space at around 2005.

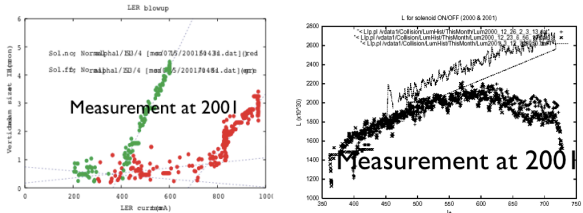

Figure 16: Beam size blow-up and luminosity limitation in 2000-2001.

Synchro-betatron sideband signal, which indicates head-tail instability caused by electron cloud, has been observed [9]. The synchro-beta signal synchronizes with the beam size blow-up: when the beam size blow-up is suppressed by the solenoid, the sideband disappears, vice versa. Figure 17 shows the betatron and synchrotron sideband spectra along the bunch train. Vertical axis is bunch train, head to tail. Betatron signal, which is left white line, shift positively. It is tune shift due to electron cloud. Right white line is a positive synchrotron sideband, whose frequency is $\nu_y+a\nu_s$, where $1<a<2$. The separation of betatron and sideband is larger than $\nu_s=0.025$, because of the fast head-tail instability; perhaps mode coupling of m=1 and m=2.

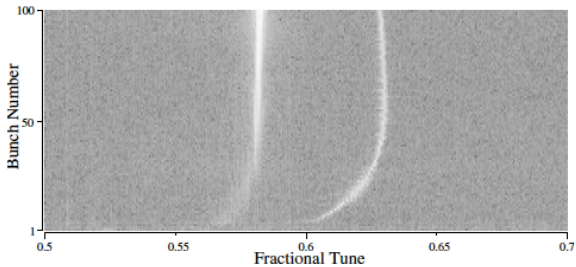

Figure 17: Measured spectrum for vertical bunch motion. Lower and upper is FFT signal of head and tail bunches [9].

Figure 18 shows simulation result of the fast head-tail instability using Eqs. (6) [11]. Beam size evolution and FFT of the bunch motion are depicted in top and bottom pictures. The figure shows the threshold electron density is around $8\times10^{11}$ m$^{-3}$. Upper synchrotron sideband appears above $10^{12}$ m$^{-3}$. The separation from betatron tune is larger than synchrotron tune 0.025 and increase for higher electron density. The result agrees well with the measurement.

The instability was not suppressed by the bunch-by-bunch feedback; suppression of the dipole mode did not affect the emittance growth due to the head-tail motion. Figure 19 shows measured betatron and sideband spectra for changing the bunch-by-bunch feed back gain. The left peak, which is betatron motion, is suppressed but the right peak was not changed. Figure 20 shows the simulation result for the feed back response. Sideband peak is not suppressed by the feedback in the simulation. The result agrees well with the measurement.

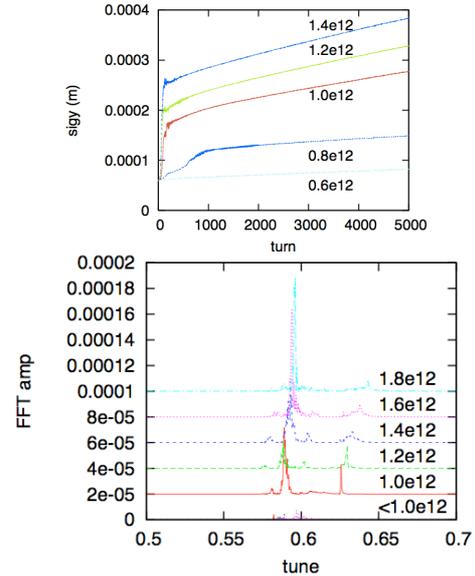

Figure 18: Simulation result of the fast head-tail instability using Eq. (6). Top picture shows beam size growth for various electron densities. Bottom picture shows FFT of the bunch motion in the top picture.

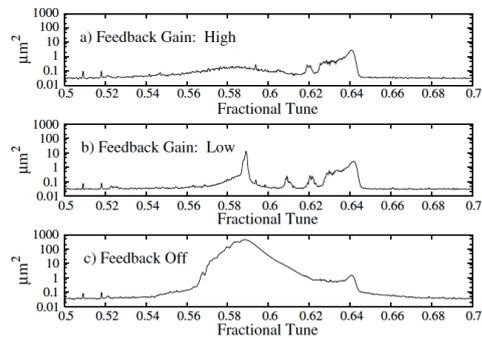

Figure 19: Averaged spectra of all bunches with feedback gain (a) high, (b) low and (c) set to zero [9]. The vertical betatron peak is 0.588 and sideband peak is around 0.64.

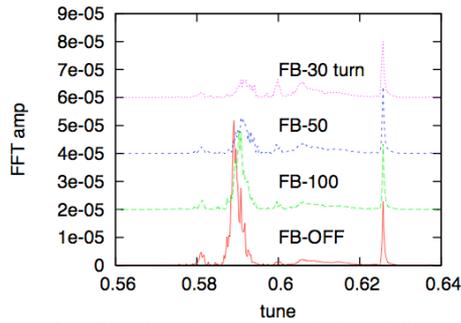

Figure 20: Synchro-betatron side band for various gain of the bunch-by-bunch feed back in the simulation, where $\rho_e=10^{12}$ m$^{-3}$.

## SUMMARY AND CONCLUSIONS

Electron cloud instability had been observed in positron storage rings, KEK-PF, BEPC-II and KEKB. The instabilities caused by electron clouds are conjectured at KEK-PF, confirmed at BEPC, and established in KEKB. The journey was the way to success of KEKB. Cooperation of experiments, theory and simulations has been satisfactory.

Mode spectra observed in the coupled bunch instability are clear evidence of the electron cloud effects. Collective motion of electrons interacting with beam is reflected in the unstable mode spectra. Electrons moving in drift space induce broad modes higher than m~h/2. Electrons in weak solenoid field induces slow and positive broad modes m~+0.

Synchro-betatron sideband spectra in the single bunch instability appear as positive sideband in KEKB. Simulations can give results consistent with the measurement.

## ACKNOWLDGMENT

The author thanks KEK-PF, BEPC and KEKB teams for fruitful discussions on the experimental results.

## REFERENCES

[1] M. Izawa, Y. Sato, T. Toyomasu, Phys. Rev. Lett. 74, 5044 (1995).
[2] T. Toyamasu et al., private communications (1995).
[3] K. Ohmi, Phys. Rev. Lett. 75, 1526 (1995).
[4] Z. Guo et al., Phys. Rev. ST-AB 5, 124403 (2002).
[5] M. Tobiyama et al., Phys. Rev. ST-AB 9, 012801 (2006).
[6] S. Win et al., Phys. Rev. ST-AB 8, 094401 (2005).
[7] H. Fukuma et al., Proceedings of HEACC2001.
[8] K. Ohmi, Z. Zimmermann, Phys. Rev. Lett. 85, 3821 (2000).
[9] J. Flanagan et al., Phys. Rev. Lett. 94, 054801 (2005).
[10] A. Chao, "Physics of Collective Beam Instabilities in High Energy Accelerators", Wiley-Interscience Publication, p61.
[11] E. Benedetto et al., Proceedings of PAC07, p. 4033 (2007).